# Time-varying microwave photonic filter for arbitrary waveform signal-to-noise ratio improvement

Dong Ma[a,b], and Yang Chen[a,b,*]

[a] *Shanghai Key Laboratory of Multidimensional Information Processing, East China Normal University, Shanghai 200241, China*
[b] *Engineering Center of SHMEC for Space Information and GNSS, East China Normal University, Shanghai 200241, China*
[*] *ychen@ce.ecnu.edu.cn*

**ABSTRACT**

A time-varying microwave photonic filter (TV-MPF) based on stimulated Brillouin scattering (SBS) is proposed and utilized to suppress the in-band noise of broadband arbitrary microwave waveforms, thereby improving the signal-to-noise ratio (SNR). The filter-controlling signal is designed according to the signal to be filtered and drives the TV-MPF so that the passband of the filter is always aligned with the frequencies of the signal to be filtered. By continuously tracking the signal spectral component, the TV-MPF only retains the spectral components of the signal and filters out the noise other than the spectral component of the signal at the current time, so as to improve the in-band SNR of the signal to be filtered. An experiment is performed. A variety of signals with different formats and in-band SNRs are used to test the noise suppression capability of the TV-MPF, and the waveform mean-square error is calculated to quantify the improvement of the signal, demonstrating the excellent adaptability of the proposed TV-MPF to different kinds of signals.

## 1. Introduction

Microwave filters play a crucial role in microwave signal processing and have been extensively studied [1,2]. A microwave photonic filter (MPF) is a photonic subsystem with unique properties such as low loss, large bandwidth, and immunity to electromagnetic interference that can carry the tasks that conventional microwave filters can or cannot carry [3]. One of the main tasks of a linear time-invariant (LTI) filter is to suppress the out-of-band noise and interference. However, most of the reported MPFs are LTI filters, which can do nothing about the in-band noise and interference, thereby severely limiting the application scenarios of MPF.

Time-varying filters (TVFs) refer to the filter whose frequency response changes with time, which can be used to suppress the in-band noise and interference so as to improve the in-band signal-to-noise ratio (SNR). Therefore, it has been widely used in signal and image processing and measurement systems [5,6]. High-frequency and high-speed tunable electrical filters are difficult to design in the electrical domain. Therefore, TVFs for electrical signals are difficult to implement in the analog domain [1,2], and they are usually designed in the digital domain. However, for microwave signals to be filtered, if the TVF is designed in the digital domain, the limited sampling rate of the analog-to-digital converter and the huge amount of data to be processed will greatly limit the real-time filtering capability of the TVF [5]. The research about MPF has focused on the reconfigurability, tunability, and high selectivity, which allows the frequency response of the MPF to be designed flexibly according to the requirements, opening up attractive possibilities for microwave signal processing [3,4]. Furthermore, the excellent properties of MPF, such as high reconfigurability and fast tunability [7-12], also make it possible to design time-varying frequency responses and perform real-time TVF directly in the optical domain.

To achieve fast tuning of the MPF center frequency, phase-modulation-to-intensity-modulation conversion is a good method [8,9]. However, in [8,9], the MPF only has a single passband though it can be scanned. In [13], we took the lead in realizing MPF with multiple time-varying passbands and applied it to a Fourier-domain mode-locked optoelectronic oscillator to generate arbitrary microwave waveforms. The ultra-fast tunability of MPF has been demonstrated, which provides a solution for us to build a real-time TV-MPF to fill the gap in the electrical domain. However, the MPF realized in [13] is implemented using a phase-shifted fiber Bragg grating, and the system is with large insertion loss. MPFs based on stimulated Brillouin scattering (SBS) provide much lower insertion loss while ensuring extremely narrow filter bandwidth, with the assistance of the ultra-narrow Brillouin gain [9-12]. We believe the SBS-based MPF is a promising solution for the processing of weak signals.

In this letter, a time-varying microwave photonic filter (TV-MPF) based on SBS is proposed for suppressing the in-band noise of broadband arbitrary microwave waveforms, thereby improving the SNR of the waveforms. An experiment is performed to verify the noise suppression capability of the proposed TV-MPF. A variety of signals with different formats and in-band SNRs are tested, and the waveform mean-square error (MSE) is calculated to quantify the improvement of the signal by the TV-MPF, demonstrating the excellent adaptability of the proposed TV-MPF to different kinds of signals.

## 2. Principle and experimental results

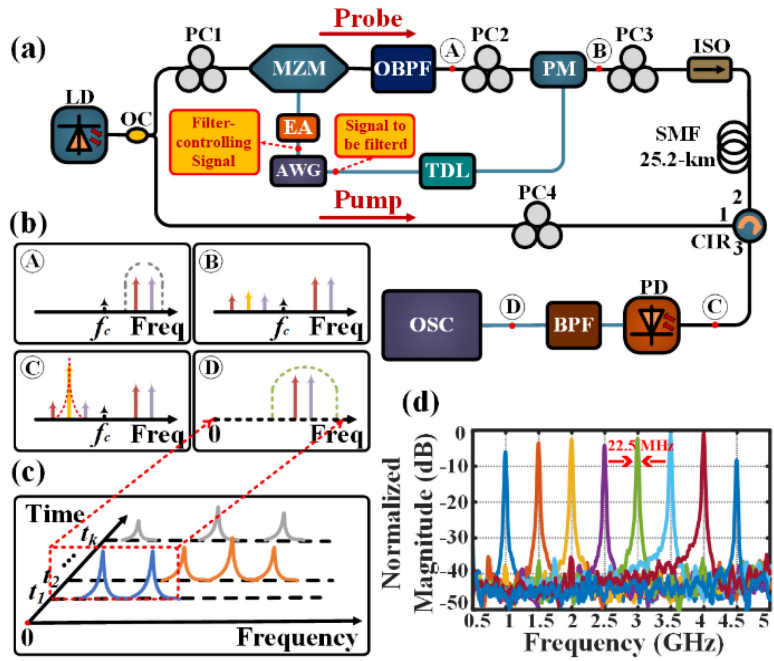

**Fig. 1.** (a) Schematic diagram of the proposed TV-MPF, (b) spectra at different positions in the system at $t_1$, (c) the time-varying response at different moments, (d) the frequency tunability of the proposed TV-MPF with a single passband from 1 to 4.5 GHz.

Fig. 1(a) illustrates the structure and experimental setup of the TV-MPF. A 15-dBm optical carrier from a laser diode (LD, ID Photonics CobriteDX1-1-C-H01-FA) with a frequency of 193.348 THz is split into two branches by a 50:50 optical coupler (OC). The light from the upper branch is sent to a null-biased Mach-Zehnder modulator (MZM, Fujitsu FTM-7938EZ) through a polarization controller (PC1). The filter-controlling signal generated from an arbitrary waveform generator (AWG, Keysight, M8195A) is applied to the RF port of the MZM via an electrical amplifier (EA, CTT CLM/145-7039-293B). The carrier-suppressed double-sideband (CS-DSB) modulated signal from the MZM is filtered by an optical bandpass filter (OBPF, EXFO, XTM-50) to select only the upper optical sideband, which is then used as the new optical carrier and injected into a phase modulator (PM, iXblue, MPZ-LN-10) through PC2. The PM is driven by the signal to be filtered via an electrical tunable delay line (TDL, Sage 6705). The signal to be filtered is also generated from the AWG, superimposed with additive Gaussian white noise, and delayed in the digital domain so that the signal to be filtered matches the control signal in time. The TDL is used to achieve higher time matching accuracy in the experiment.

The optical signal from the PM is injected into a section of 25.2-km single-mode fiber (SMF) through an isolator as the probe wave. In the lower branch of the OC output, the optical carrier used as the pump wave is reversely launched into the SMF, where it interacts with the counterpropagating probe wave from the upper branch. Among the optical sidebands generated by phase modulation, only the sidebands falling into the Brillouin gain spectrum (BGS) are amplified, as shown in the spectrum at point C in Fig. 1(b). The optical signal is detected in a photodetector (PD, Nortel PP-10G) to generate the signal after filtering, which is then filtered by a band-pass filter (BPF, KGL YA351-2, 2.4-4.0 GHz) to remove the out-of-band noise and spurious interference. The electrical spectrum from the PD is shown at point D in Fig. 1(b), which is the same as that of the signal to be filtered but with less in-band noise. The waveform is monitored by an oscilloscope (OSC, R&S RTO2032). It should be mentioned that four PCs are used in the system. PC1 and PC2 are used to ensure the maximum MZM and PM modulation efficiency, while PC3 and PC4 are used to adjust the polarization states of the pump light and probe light to ensure the most efficient SBS interaction. The above system maps the optical filter formed by the BGS to an electrical filter whose center frequency is controlled by the filter-controlling signal. The frequency components of the filter-controlling signal change with time, which leads to time-varying frequency response as shown in Fig. 1(c).

As shown in Fig. 1, the signal to be filtered is a broadband microwave signal with superimposed noise, while the filter-controlling signal is designed according to the signal to be filtered. Commonly, the signal to be filtered is with low SNR and the filter-controlling signal is a copy of the signal but with good SNR. Therefore, we need to have exact knowledge of the signal to be filtered when using the proposed TV-MPF. Indeed, this is the actual situation in many application scenarios. For example, in radar detection systems, we can use the transmitted signal having the same format as the echo signal as the filter-controlling signal to improve the signal quality of the weak echo signals. The filter-controlling signal drives the TV-MPF so that the passband of the filter is always aligned with the frequencies of the signal to be filtered. By continuously tracking the signal spectral component, the TV-MPF only retains the spectral components of the signal and filters out the noise other than the spectral component of the signal at the current time, so as to improve the in-band SNR of the signal to be filtered.

In the work, the noise suppression capability of the proposed TV-MPF is quantified by the waveform MSE. Taking the response of the system and the OSC into consideration, the reference waveform is generated from the PD by sending a signal to the PM without adding additional noise. The waveform from the PD when the same signal with additional noise is sent to the PM is compared with the reference waveform to obtain the waveform MSE. The noise suppression by the TV-MPF can be obtained by comparing the MSE before and after the TV-MPF.

Firstly, the tunability of the TV-MPF is measured. A single-tone signal is used as the filter-controlling signal and sent to the MZM, while the sweep output of a vector network analyzer (VNA, Agilent 8720ES) is connected to the PM and the output from the PD is connected to the VNA input. The tunability of the filter is verified by adjusting the frequency of the filter-controlling signal. In Fig. 1(b), the center frequency of the MPF equals the Brillouin frequency shift (BFS) plus the frequency of the filter-controlling signal because the positive sidebands from the MZM are selected by the OBPF. In the experiment, the negative sidebands are selected so that the center frequency of the MPF equals to the absolute value of the BFS minus the frequency of the filter-controlling signal. In the following results, we also show the results within 4 GHz. Fig. 1(d) shows the frequency response of the TV-MPF when the signal frequency applied to the MZM changes from 11.8 to 15.3 GHz. As can be seen, the center frequency of the passband varies from 1.0 GHz to 4.5 GHz and the 3-dB bandwidth is measured to be around 22.5 MHz.

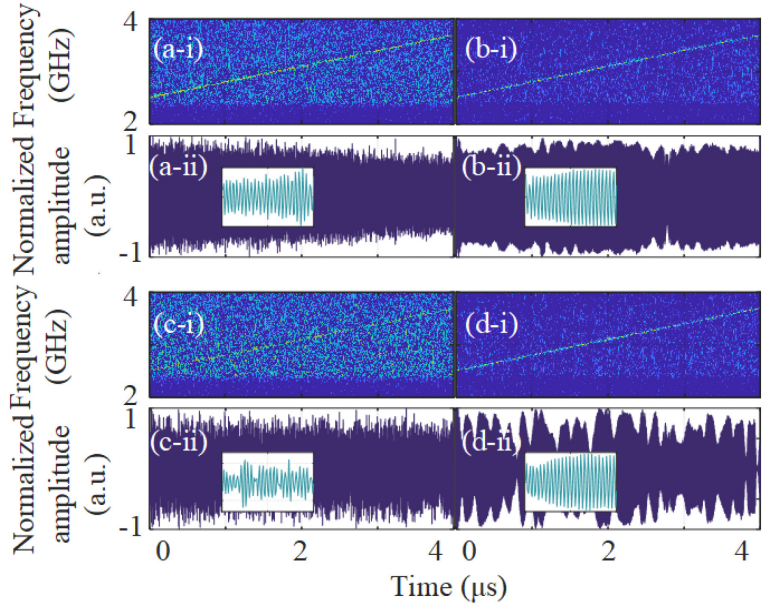

**Fig. 2.** Comparison of LFM signals with superimposed noise before and after the TV-MPF. Waveforms and TFDs with 8-dB in-band SNR (a) before and (b) after the TV-MPF, with -4.5-dB in-band SNR (c) before and (d) after the TV-MPF. (i) TFDs, (ii) waveforms. The insets show the corresponding waveforms from 0-10 ns.

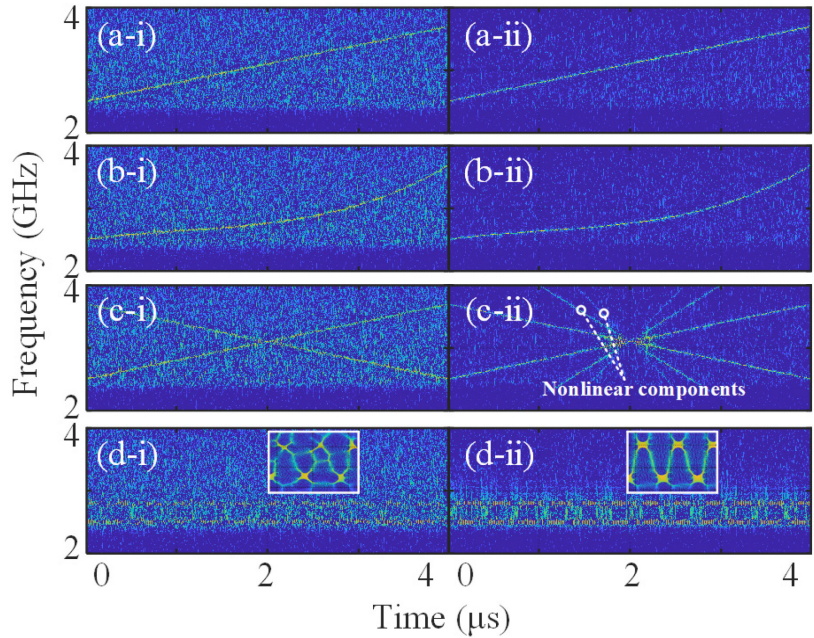

**Fig. 3.** TFDs of (a) LFM signals, (b) NLFM signals, (c) DLFM signals, and (d) FH signals (i) before and (ii) after the TV-MPF. The insets show the TFDs of the FH signals from 0-50 ns.

Secondly, linearly frequency-modulated (LFM) signals with a bandwidth of 1.2 GHz, a center frequency of 3.1 GHz, and different SNRs are used as the signal to be filtered to demonstrate the real-time noise suppression capability of the TV-MPF. The in-band noise examined here is the noise that cannot be filtered by the BPF (2.4-4.0 GHz), thus the noise bandwidth is 1.6 GHz. The filter-controlling signal is an LFM signal with a higher center frequency of 13.9 GHz. When the in-band SNR is 8 dB, the waveform and time-frequency diagram (TFD) before filtering are shown in Fig. 2(a). The noise distribution can be seen from the TFD, in which the noise out of the passband of the BPF is well suppressed (2-2.4 GHz in the figure). However, the in-band noise is still strong, which makes the waveform still distorted compared with the noise-free signal, with a waveform MSE of 0.1974. After filtering using the TV-MPF, the in-band noise shown in Fig. 2(b-i) is significantly reduced. The waveform is also improved, as can be seen in Fig. 2(b-ii), and the waveform MSE is reduced to 0.1575. When the in-band SNR is further reduced to -4.5 dB, the waveform and TFD before and after filtering are shown in Fig. 2(c) and (d). As the decrease of the SNR, the waveform distortion is more serious, with a waveform MSE of 0.2779. After filtering, the in-band noise is greatly reduced and the waveform MSE is also reduced to 0.1847. The above comparison shows that the in-band noise can be suppressed to a certain degree after the TV-MPF and the signal waveform can be effectively improved by TV-MPF, within a certain in-band SNR range.

Then, the TV-MPF is further verified by using different kinds of broadband frequency-agile signals. In this study, the in-band SNR of the signal to be filtered is fixed at 3 dB. Fig. 3(a)-(d) show the TFDs of LFM signals, non-linearly frequency-modulated (NLFM) signals, dual-chirp LFM (DLFM) signals, and frequency-hopping (FH) signals before and after the TV-MPF. The LFM, NLFM, and DLFM signals have a bandwidth of 1.2 GHz, a center frequency of 3.1 GHz, and a period of 4 us. The FH signal has two frequencies at 2.5 and 2.8 GHz, and the hopping period is 10 ns. The corresponding controlling signal is a copy of the signal to be filtered after a frequency shifting of 10.8 GHz. For signals like LFM, NLFM, and FH signals, which have only one frequency at a specific time, the TV-MPF can optimize the SNR significantly from the figures. For DLFM, which has multiple frequencies at a specific time, the SNR is also greatly improved as well as introducing some weak harmonic frequencies. The generation of the weak harmonic frequencies is due to the limitation in our experiment: the saturation power of the EA is low, resulting in the distortion of the filter-controlling signal after EA1. Therefore, the weak harmonic frequencies in Fig. 3(c-ii) can be eliminated by using a non-distorted filter-controlling signal.

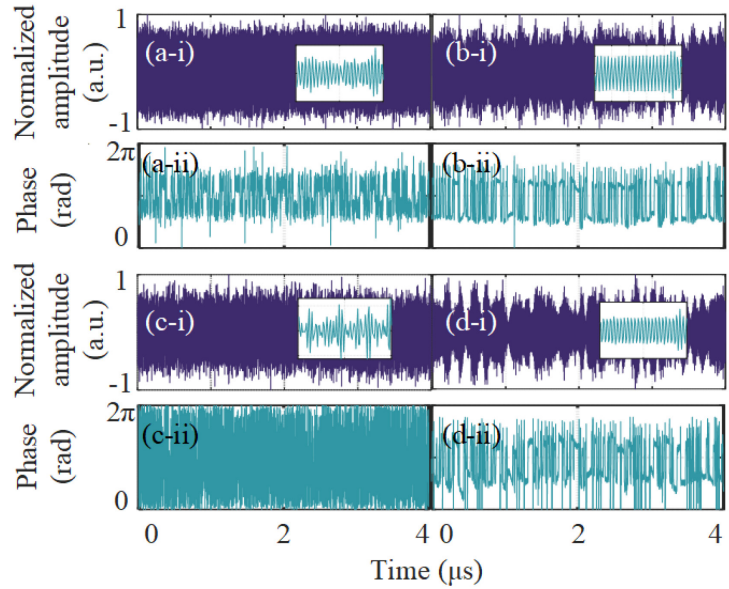

**Fig. 4.** Recovered phase information and waveforms for binary phase-coded signals with 8-dB in-band SNR (a) before and (b) after the TV-MPF, with -2-dB in-band SNR (c) before and (d) after the TV-MPF. (i) Waveforms, (ii) recovered phase information. The insets show the corresponding waveforms from 0-10 ns.

The proposed TV-MPF can also perform time-varying filtering for more complex signals. Then, the noise reduction for binary phase-coded signals is further demonstrated using the TV-MPF. The phase-coded signal is set to have 400 bits in a period of 4 us. The carrier frequency of the phase-coded signal is 2.5 GHz. Coded in the same way, the controlling signal in the experiment is centered at 13.3 GHz, considering the 10.8-GHz BFS. The waveforms and corresponding recovered phase information before and after the TV-MPF with in-band SNRs of 8 and -2 dB are shown in Fig. 4. As can be seen, when the SNR is 8 dB, we can observe discrete phases. However, due to the large in-band noise, the phase information can be hardly recovered. After filtering using the TV-MPF, as shown in Fig. 4(b-i), all phase information is recovered correctly, and π phase jumps can be observed. When the SNR is reduced to -2 dB, the waveform is seriously distorted and cannot recover any phase information. After the TV-MPF, the phase information is recovered, though it is not as good as that when the SNR is 8 dB.

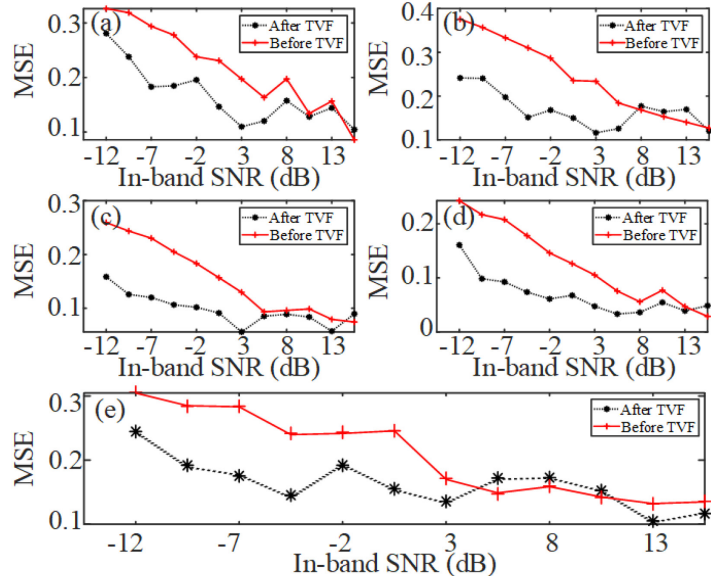

**Fig. 5.** Waveform MSEs versus in-band SNRs for (a) LFM signal, (b) NLFM signal, (c) DLFM signal, (d) FH signal, and (e) phase-coded signal, before and after TV-MPF filtering.

To further investigate the noise suppression capability of the TV-MPF for different kinds of signals, the waveform MSEs of different formats before and after the TV-MPF are verified. Fig. 5 shows the waveform MSEs of LFM signals, NLFM signals, DLFM signals, FH signals, and phase-coded signals before and after filtering using the TV-MPF. The in-band SNR is adjusted from -12 to 15.5 dB. As can be seen, when the in-band SNR is below 5.5 dB, all kinds of signal waveforms can be improved to a certain degree after the TV-MPF. In comparison, when the signal has good SNR, the effect of TV-MPF on signal in-band noise suppression is not obvious. Although the noise outside the BGS at the current time can be filtered out by the time-varying filtering enabled by SBS, the noise in the narrow BGS will be amplified. When the SNR is very high, the additional noise introduced by amplifying the noise inside the BGS exceeds the noise filtered out outside the BGS, leading to no SNR improvement under a very good SNR. Therefore, the TV-MPF should be used in the condition with low SNR, and the suppression of the in-band noise of signals with low SNR is exactly what the real-world system needs. For example, the proposed TV-MPF can be used to improve the low-SNR radar echo signals.

Considering the operating bandwidth of the equipment and devices in our laboratory, especially that of the OSC, the signal to be filtered is selected to be lower than 4 GHz for convenience. As discussed above, the negative sidebands are selected so that the center frequency of the MPF equals the absolute value of the BFS minus the frequency of the filter-controlling signal. In the experiment, we also verify the TV-MPF when the positive sidebands are selected by the OBPF as shown in Fig. 1. In this case, the center frequency of the MPF equals to the BFS plus the frequency of the filter-controlling signal. The signal to be filtered is set as an LFM signal from 12.3 to 12.8 GHz with an in-band SNR of 10 dB, and the filter-controlling signal is set as an LFM signal from 2.5 to 3 GHz. The EA and PD are replaced by devices that can be operated in the corresponding frequency bands (EA, TELEDYNE COUGAR AR3069B; PD, Discovery Semiconductor DSC-40S). The high-frequency signal output from the PD is down-converted to 1.5 to 2 GHz using a mixer (M/A-COM, M14A), and then sampled by the OSC. The waveform MSE is calculated to be 0.1375, which is much improved compared with the waveform MSE of 0.2022 without TV-MPF filtering. Therefore, the TV-MPF can also provide effective waveform improvement for high-frequency signals.

## 3. Conclusions

In summary, we have proposed a TV-MPF based on SBS and applied it successfully to suppress the in-band noise of broadband arbitrary microwave waveforms in real-time. To the best of our knowledge, this is the first demonstration of improving arbitrary waveform signal SNR with the assistance of microwave photonics. Thanks to the narrow BGS provided by the SBS effect and the flexible filter-controlling signal associated with the signal to be filtered, the filtering response of the TV-MPF changes in real-time with the signal to be filtered, so as to filter out the noise other than the spectral component of the signal at the current time. An experiment is performed. The signal quality improvement of LFM signals, NLFM signals, DLFM signals, FH signals, and phase-coded signals are demonstrated, and the improvement is quantified by waveform MSE. The proposed TV-MPF will provide a feasible solution for improving the SNR of weak signals.

**Funding**

National Natural Science Foundation of China (NSFC) (61971193); Natural Science Foundation of Shanghai (20ZR1416100); Science and Technology Commission of Shanghai Municipality (18DZ2270800).

**Conflicts of interest**

The authors declare no conflicts of interest.